\documentclass[10pt,conference]{IEEEtran}

\usepackage{cite}
\usepackage{graphicx}
\usepackage{amsmath}
\usepackage{amssymb}
\usepackage{amsfonts}
\usepackage{algorithmic}
\usepackage{braket}
\usepackage{url}
\usepackage{caption}
\usepackage{subcaption}

\title{QSweep: Pulse-Optimal Single-Qudit Synthesis}

\author{
    \IEEEauthorblockN{Ed Younis}
    \IEEEauthorblockA{
        Computational Science Department\\
        Computational Research Division\\
        Lawrence Berkeley National Laboratory\\
        Berkeley, CA, 94720, USA\\
    }
\and
    \IEEEauthorblockN{Noah Goss}
    \IEEEauthorblockA{
        Quantum Nanoelectronics Laboratory\\
        Department of Physics\\
        University of California at Berkeley\\
        Berkeley, CA, 94720, USA\\
    }
}

\begin{document}

\maketitle


\begin{abstract}
The synthesis of single-qudit unitaries has mainly been understudied, resulting in inflexible and non-optimal analytical solutions, as well as inefficient and impractical numerical solutions. To address this challenge, we introduce QSweep, a guided numerical synthesizer that produces pulse-optimal single-qudit decompositions for any subspace gateset, outperforming all prior solutions. When decomposing ququart gates, QSweep created circuits 4100x (up to 23500x) faster than QSearch with an average of 7.9 fewer pulses than analytical solutions, resulting in an overall 1.54x and 2.36x improvement in experimental single-qutrit and ququart gate fidelity as measured by randomized benchmarking.
\end{abstract}

\section{Introduction}
\label{introduction}

Qudits, quantum units of information extending beyond the binary domain, have been experimentally demonstrated on trapped ion~\cite{ringbauer2022universal}, photonic~\cite{malik2016multi,reck1994experimental}, neutral-atom~\cite{anderson2015accurate}, and superconducting~\cite{goss2022high,qutritrb} quantum machines with a growing interest due to their superior computational capacity. Algorithm designers have effectively utilized qudits in simulation~\cite{gustafson2022noise}, optimization~\cite{bottrill2023exploring}, and cryptography~\cite{groblacher2006experimental} applications, among many others. All show that quantum programs effectively leveraging qudits can find asymptotic improvements in performance~\cite{gokhale2019asymptotic}. Consequently, there has been a growing development of compilers that seek to take advantage of these resources automatically~\cite{litteken2023qompress}.

While the development of qudit compilation tools is expanding, the focus is primarily on two-qudit instructions, often completely ignoring the single-qudit gates. This mindset, adopted from qubit-based compiler development, presents a significant problem in the qudit space: single-qudit gate decompositions grow quadratically with radix. For qubits, an arbitrary single-qubit rotation typically requires five native instructions with some variance depending on a machine's instruction set. Moreover, machines execute these instructions with very high fidelity. However, higher-dimensional single-qudit instructions are much more error-prone, making it critical to implement these gates efficiently. Yet, they are not in practice. State-of-the-art decompositions break down qutrit, three-level qudits, into 15 native gates and ququarts, four-level qudits, into 30 native gates~\cite{de2018simple}. These decompositions can quickly grow to dominate circuit complexity and runtime.

Numerical synthesizers~\cite{qsearch} have been effectively used to find optimal circuits in the qubit space with respect to gate counts. Applying existing techniques to qudit decompositions is possible but can take several hours to find optimal solutions, whereas analytical methods find sub-optimal solutions in fractions of a second. This runtime penalty ensures that these tools are not used in practice, as the problem of single-qudit decompositions must be performed hundreds to thousands of times for a single experiment or algorithm execution. Furthermore, these synthesizers typically do not distinguish between native gates, leading to a non-physically motivated definition of optimality. It is often the case that only some native gates are physically implemented with pulses on quantum machines, while others are purely virtual. Minimizing the number of physical pulses is more critical than the total gate count. We call this measure pulse-optimality.

To address these challenges efficiently and effectively, we make the following contributions:
\begin{itemize}
    \item We modify a pre-existing analytical solution to prioritize pulses in less error-prone subspaces, increasing fidelity.
    \item Based on this modified algorithm, we introduce QSweep, a guided numerical synthesizer that produces pulse-optimal single-qudit decompositions for any user-configurable subspace gateset, outperforming all prior solutions. It matches the performance of analytical solutions and improves the quality of numerical synthesizers.
    \item We experimentally demonstrate the advantage of QSweep with single-qutrit and ququart randomized benchmarking on a superconducting quantum device, leading to a $1.54\times$ and $2.36\times$ improvement in fidelity for qutrit and ququart instructions, respectively.
\end{itemize}

The following section provides relevant background on qudit quantum computing, synthesis, and benchmarking. Section~\ref{related} surveys the field of qudit synthesis and decomposition tools, leading up to the description of our analytical and numerical algorithms in Section~\ref{algorithm}. We then evaluate our tools in Section~\ref{evaluation} and conclude in Section~\ref{conclusion}.
\section{Background}
\label{background}
\subsection{Qudit-based Computation}\label{sec:qudit-comp}
Qudits are the $d$-level generalization of qubits, where formally a general pure qudit state can be written as $\ket{\psi} = \sum_{i=0}^{d-1}\alpha_{i}\ket{i}$, with $\alpha_i \in \mathbb{C}$ subject to the constraint $\sum_{i=0}^{d-1} |\alpha_{i}|^2 = 1$. Notably, the effective Hilbert space available for qudit based quantum computation grows as $d^N$, where $N$ is the effective number of entanglable qudits in a quantum processor, meaning techniques that can enable larger qudit dimension $d$ can in principle translate to large increases in computational power for the same number of qunatum units (qudits) in a quantum processor. 

One can generalize the single-qubit Pauli group to $d$ dimensions as the so-called Weyl-Heisenberg basis $\mathbb{W}_d = \{W_{x,z} = X^x Z^z, x,z \in \mathbb{Z}_d \}$. In the Weyl-Heisenberg basis, the generalized $X$ and $Z$ operators are defined by their action on the basis state $\ket{n}$ as $X\ket{n} = \ket{n \oplus_d 1}$ and $ Z\ket{n} = \omega^n \ket{n}$, where $\omega = e^{i 2\pi/d}$ is the $d$-th root of unity. To generalize the Cliffords, we first must note that the Weyl-Heisenberg basis is not closed under multiplication, and therefore not a proper group. We therefore define, the ``extended" Weyl-Heisenberg group as $\mathbb{E}\mathbb{W}_d = \mathbb{U}(1) \mathbb{W}_d$. With this, we can define the qudit-Clifford group as the set of operators which normalize the extended Weyl-Heisenberg group, i.e. $\mathbb{C}_d = \{ U \in \mathbb{U}(d) :  U \mathbb{E}\mathbb{W}_d U^\dag = \mathbb{E}\mathbb{W}_d \}$ \cite{bengtsson2017discrete}.

\subsection{Circuit Synthesis}

Quantum programs are expressed as circuits where the wires extend from left to right through time and represent the qudits. Gates reflect operations applied to the qudit's state that they touch. Unitary operators can mathematically represent all individual gates and whole circuits. While there are infinitely many quantum gates, machines only support a small subset of instructions called their native gateset. In quantum computing, circuit synthesis is the process of translating a high-level description of a program, often a large unitary matrix, into an executable circuit for a given quantum machine.

\subsection{Randomized Benchmarking}

Randomized Benchmarking (RB) \cite{PhysRevA.77.012307} is a technique used to quantify the average error rates associated with the Clifford gates on a quantum processor. RB consists of generating random Clifford circuits of varying depths, where at each depth the Clifford gates are sampled uniformly. The final gate in each RB circuit is chosen to inverse all prior operations, such that the entire circuit decomposes to the identity, allowing easy comparison between experimentally measured and ideal results. The generation and measurement of these random circuits is repeated many times at each selected depth, yielding robust experimental results for the average performance of the quantum processor. By fitting an exponential decay curve to the extracted average fidelity vs. depth, an estimate for the average fidelity of each Clifford gate is obtained that is decoupled from the state preparation and measurement errors (SPAM) present in the device. As the Clifford group can be generalized to qudits (see sec.\ref{sec:qudit-comp}), the RB procedure can be readily applied on a qudit processor, and has been performed for experimentally for qudits of $d=3$~\cite{qutritrb} and $d=4$~\cite{PhysRevX.13.021028}. 
\section{Related Works}
\label{related}
Unitary factorization and quantum circuit synthesis are closely related problems, with the main difference being the desired outcome. The former seeks to express a complex unitary operation as a product of simpler ones. Meanwhile, the latter aims to implement a target operation with a quantum circuit using a fixed gateset. While there are subtle differences generally, they are equal for single-qudits with two main approaches: bottom-up and top-down. This section will explore the single-qudit solutions designed and utilized in practice.

\subsection{Top-Down Methods}

Top-down methods use fixed, algebraic rules to break down a target unitary into smaller ones while maintaining equality between the target operation and the current set of unitaries or gates. This process is often recursive or hierarchical. Top-down approaches for the single-qudit problem typically follow from unitary parameterization schemes. Practitioners in \cite{qutritrb} adapted the method in \cite{dita2003factorization} to decompose arbitrary qutrit rotations. They also renumbered the qudit levels during synthesis to avoid overuse of the higher levels. This method generalizes to higher-radix qudits but always produces a result with the maximum number of pulses.

\subsection{Bottom-Up Methods}

Bottom-up approaches continuously apply small unitaries or gates to build up to the target operation. There are both analytical and numerical solutions that follow this paradigm. Analytical solutions follow a fixed set of pre-determined procedures to build a structured result, whereas numerical ones employ continuous unitary parameterization and numerical optimization to solve for parameters. 

\subsubsection{Elimination Methods}

\begin{figure*}[t!]
    \centering
    \begin{subfigure}[b]{0.3\textwidth}
        \centering
        \includegraphics[width=\textwidth]{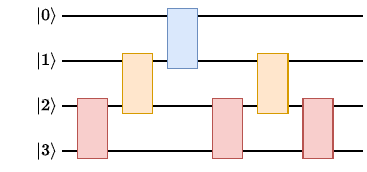}
        \caption{Column-by-Column}
        \label{fig:cbc}
    \end{subfigure}
    \begin{subfigure}[b]{0.3\textwidth}
        \centering
        \includegraphics[width=\textwidth]{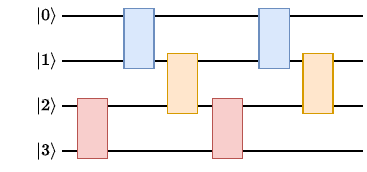}
        \caption{Square}
        \label{fig:square}
    \end{subfigure}
    \begin{subfigure}[b]{0.3\textwidth}
        \centering
        \includegraphics[width=\textwidth]{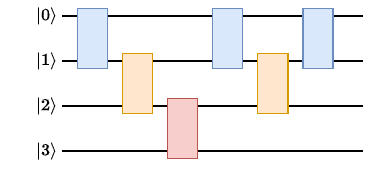}
        \caption{Row-by-Row}
        \label{fig:rbr}
    \end{subfigure}
    \caption{
        \footnotesize \it There are various valid elimination patterns to zero elements in the target matrix, each resulting in a different interaction architecture in the synthesized circuit. Figure (a) shows the column-by-column method from~\cite{de2018simple}, while (b) illustrates the square method from~\cite{clements2016optimal}. Lastly, (c) demonstrates the resulting architecture of our row-by-row method. The awkward spacing between interactions on non-overlapping subspaces is intentional, as these gates are performed sequentially in practice.
    }
    \label{fig:architectures}
\end{figure*}

Analytical bottom-up methods follow an elimination pattern and are widely used in quantum photonic architecture design. Whenever a small unitary or gate is applied, it zeroes out an element in the target unitary. Reck et al.~\cite{reck1994experimental} pioneered this approach by introducing a technique to zero out target matrix elements by applying a precisely crafted, embedded U(2) operator. Each application corresponds to a beam splitter coupling two neighboring modes in the architecture design. This process continues until the target becomes the identity. In \cite{de2018simple}, the authors introduce a zero-ordering that iterates column-by-column, resulting in a triangular architecture that interacts the higher levels many more times than the lower ones. Clements et al.~\cite{clements2016optimal} eliminates elements from the target matrix bidirectionally, producing a square design where neighboring levels interact an equal amount of times. Figure~\ref{fig:architectures} illustrates the different patterns. While these methods lead to differences in the interaction patterns of the qudit levels, they all produce architectures or circuits with the same number of gates. When applied to single-qudit synthesis, they do not account for easier inputs and consistently make circuits with the maximum length.

\subsubsection{Numerical Approaches}

QSearch~\cite{qsearch} is the canonical bottom-up numerical synthesis algorithm. Initially designed for few-qubit systems, it has recently been adapted for multi-qudit and single-qudit decomposition problems. The algorithm combines numerical optimization for parameter instantiation with a search over circuit structures, leading to near-optimal circuits. This approach works for any target gateset as it is purely numerical. The performance and scalability of QSearch depend upon the size of the synthesized system and the search's branching factor. To keep the branching factor low, QSearch groups arbitrary single-qudit rotations with two-qudit gates during circuit structure expansion. This strategy is not possible for single-qudit synthesis but also not necessary for binary decompositions as the branching factor and system size are low, targeting single-qubit gatesets with only two unique gates. However, as the radix grows, the branching factor and system size also increase, making this strategy an inefficient solution for the general single-qudit decomposition problem.
\section{Synthesis Algorithm}
\label{algorithm}

We first modify the column-by-column analytical decomposition presented in \cite{de2018simple} to a similar row-by-row elimination method. This change flips the resulting level coupling architecture, minimizing the number of interactions between the higher levels of the qudit. In practice, this subtle change is essential to maximize fidelity as the higher-level pulses are more prone to error. We then develop a guided numerical synthesis algorithm using parameter instantiation following the same row-by-row elimination method. The numerical algorithm has the added benefit of working with any gateset and using fewer pulses.

\begin{figure}[t!]
    \centering
    \includegraphics[width=\columnwidth]{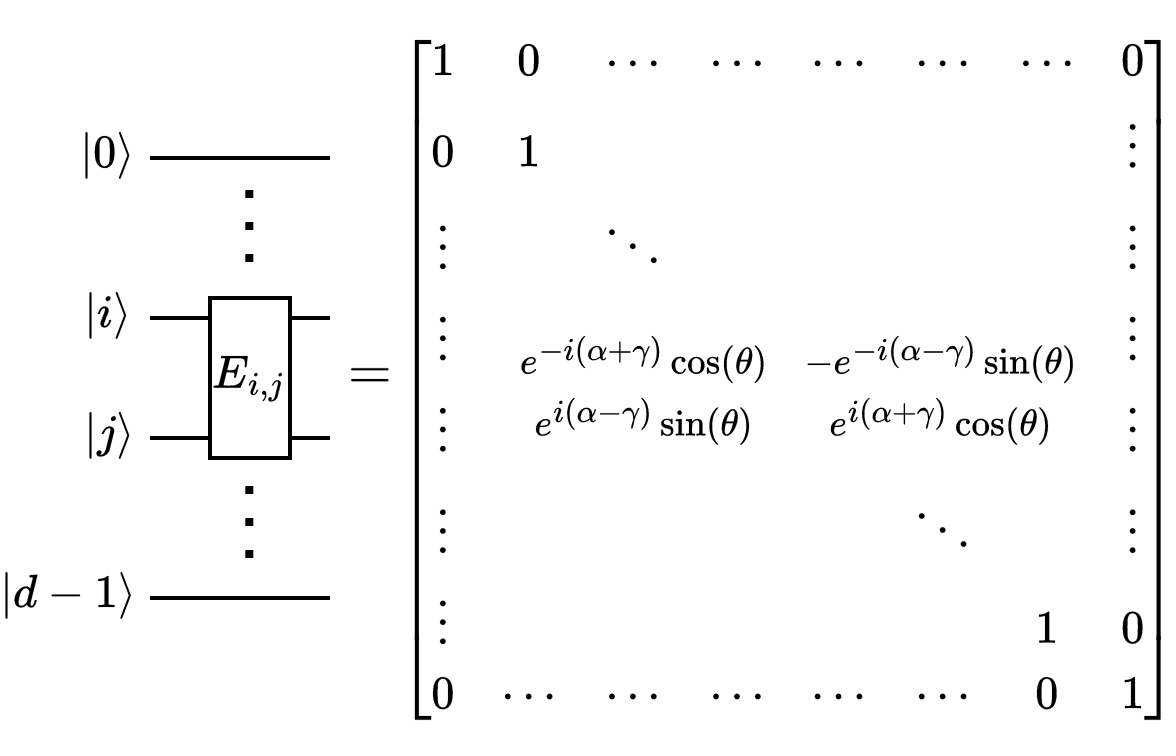}
    \caption{
        \footnotesize \it A diagrammatic interaction between the $i$-th and $j$-th modes is represented by an identity matrix with a three-parameter $U(2)$ embedded in the $i$-th and $j$-th rows and columns.
    }
    \label{fig:matrix}
\end{figure}

\begin{figure}[t!]
    \centering
    \includegraphics[width=\columnwidth]{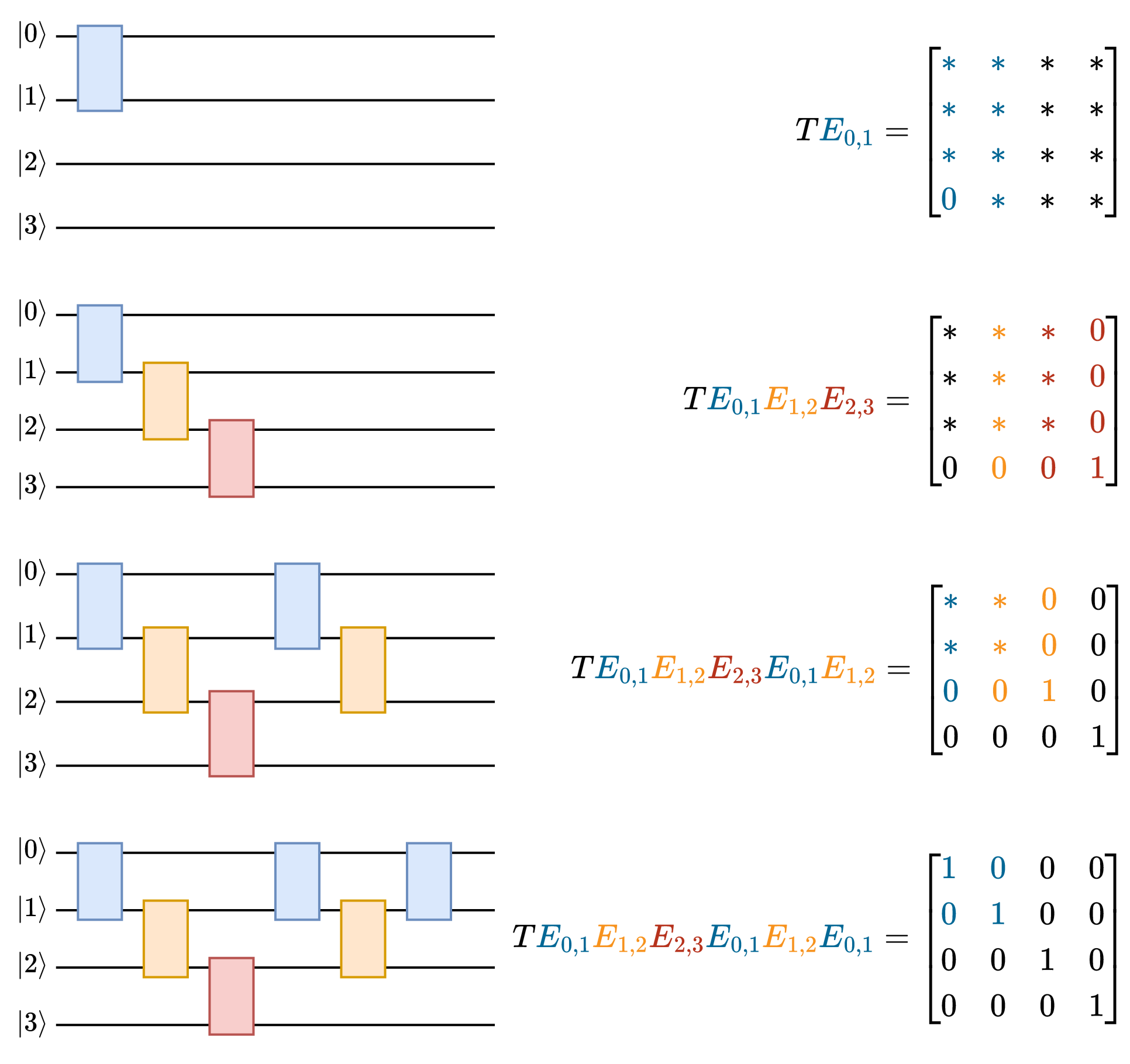}
    \caption{
        \footnotesize \it The row-by-row method eliminates elements in the target matrix one row at a time. Every right multiplication zeros an additional element. When the row is complete, the column is also zero-ed since the target and factors are unitary. Further multiplications do not affect a row with all zeros. This process builds a circuit out of the corresponding gates.
    }
    \label{fig:rbr-alg}
\end{figure}

\subsection{Row-By-Row Elimination}

Our analytical method breaks down a target unitary into a product of $E_{i,j}$ matrices, representing embedded $U(2)$ operations in a qudit's subspace. Figure~\ref{fig:matrix} depicts these factors element-wise and equates them to a circuit component, illustrating that they can be quickly converted into quantum processor instructions. As a result, the problem of factoring a unitary matrix into $E_{i,j}$ matrices is equivalent to the single-qudit synthesis problem.

We build up our solution bottom-up by incrementally appending matrix factors until we have a final result that implements the target matrix. Each additional factor zeros an element in the target matrix. What distinguishes our analytical algorithm from previous ones is the order in which we zero elements. Starting with the leftmost element in the bottom row of the matrix, we move right until we reach the diagonal. Once we have put a zero in all but the last elements of the bottom row, we will also have indirectly put zeros in the corresponding elements in the rightmost column because the target is unitary. Since we use three-parameter embedded operations, the diagonal element will also become one. Moreover, applying additional matrix factors will no longer affect the elements in the last row. Subsequently, the algorithm recurses on the smaller submatrix. Figure~\ref{fig:rbr-alg} illustrates this algorithm step by step for a ququart decomposition.

We now describe how to calculate the next matrix factor. Let $x$ be the element in the remaining target matrix that we wish to zero, and let $y$ be the next element to the right. Let $a_0, a_1, \cdots, a_{d-3}$ be the other elements in the same row. Calculate:
\begin{equation}
    s = \sqrt{1 - \sum{|a_i|^2}}
\end{equation}
then set:
\begin{equation}
    \alpha = \text{phase}\Big(\frac{y}{s}\Big) - \text{phase}\Big(\frac{-x}{s}\Big)
\end{equation}
\begin{equation}
    \gamma = \text{phase}\Big(\frac{y}{s}\Big) + \text{phase}\Big(\frac{-x}{s}\Big)
\end{equation}
\begin{equation}
    \beta = -2\arccos{\Big|\frac{y}{s}\Big|}
\end{equation}
where $\alpha, \beta, \gamma$ are the three parameters to a $E_{i,j}$ matrix as in Figure~\ref{fig:matrix}. This gives us the parameterization of the subspace operation to zero a specific element. If the element we wish to zero is in column $j$, then we embed this operation in the subspace spanned by $\ket{j}$, $\ket{j+1}$. We apply the calculated matrix factor on the right of the target to update and get the new target matrix. At the end of the algorithm, the first diagonal element may have a non-zero phase. This is not a problem, as it can be trivially solved either by applying an additional phase gate or by enforcing the input target to be a special unitary. Lastly, we improve upon previous algorithms by simply skipping elements that are already zero, making our analytical decomposition adaptive to different inputs and improving the quality of its results.

\subsection{Numerical Synthesis}

The QSweep algorithm, a numerical variant of the row-by-row algorithm, combines the benefits of the analytical algorithm and numerical synthesis algorithms without the disadvantages. While competitive in performance, it produces much shorter circuits than analytical decompositions. Additionally, it is easily configurable to different gatesets, enabling portability to all current and future qudit processors. QSweep takes as input a unitary matrix describing a $d$-level single-qudit operation and a subspace gateset. The subspace gateset specifies the target operations between neighboring levels. For example, a typical superconducting quantum computer allows arbitrary angle Z-rotations and $\sqrt{X}$ gates in each subspace.

At a high level, the algorithm proceeds the same way as the analytical row-by-row decomposition. We eliminate elements from the target matrix row-by-row until the target becomes the identity, at which point, the algorithm terminates. The main difference is that we zero elements numerically. Like QSearch, we search over the circuit structures spanned by the target gateset, performing numerical instantiation to find the parameters of each candidate circuit. Unlike Qsearch, our goal is to find parameters that best zero a specified component. Additionally, we bound our search only to append gates in the corresponding subspace. On every successful elimination, we move to the next subspace, drastically reducing the size of our search space. It is important to note that when we zero the last element in a row, we add another term to our cost function to force the optimization to calculate a one in the diagonal. This is to avoid phase corrections with additional gates.

This algorithm provides two significant advantages over the analytical variant. Firstly, the target gateset is entirely configurable, removing the need for additional, potentially costly synthesis or decomposition algorithms. Secondly, and more importantly, the numerical optimization constantly re-instantiates parameters to all circuit gates. Dynamically reconfiguring previously placed gates enables future eliminations with fewer gates. This pivotal idea leads QSweep to produce shorter circuits requiring fewer pulses.

This concept is not unique to QSweep. QSearch and other search-based numerical syntheses consistently reinstantiate parameters during circuit search. However, all previous numerical algorithms tackle the entire target matrix all-at-once. This approach causes the branching factor to grow quadratically with radix, resulting in the runtime growing exponentially. QSweep, on the other hand, breaks the synthesis problem down into small, manageable goals, which keeps the search bounded. In this regard, we call QSweep a guided algorithm.
\section{Evaluation}
\label{evaluation}

We implemented the column-by-column, row-by-row, and QSweep algorithms using Python 3.11.4 and version 1.1.1 of the BQSKit framework~\cite{bqskit}. All algorithms were evaluated on a single-core of an Epyc 7702p system. When comparing against QSearch, we utilized the BQSKit-provided implementation with recommended settings for single-qudit synthesis. All algorithms, experiment code, collected data, and circuits are available at \url{https://github.com/edyounis/QSweep}. We refer to circuits or synthesis ``quality'' as the number of pulses required in the decomposition on a device that implements virtual Z-gates with subspace $\sqrt{X}$ gates as pulses. Superstaq~\cite{superstaq} can also competitively compile qutrit and ququart unitaries for these devices; however, we couldn't perform a complete quantitative evaluation since it is proprietary.

The analytical algorithms are exact synthesizers, but both QSweep and QSearch allow a configurable error tolerance. For QSweep, we used $10^{-8}$, which produces precise results. Every result was verified to be within floating point precision, i.e., the same level of precision as the analytical algorithms. QSearch, on the other hand, is set to a higher threshold by default, leading to distances rising to $10^{-4}$ in rare cases.


\subsection{Random Unitaries}

\begin{figure*}[t]
    \centering
    \includegraphics[width=\columnwidth]{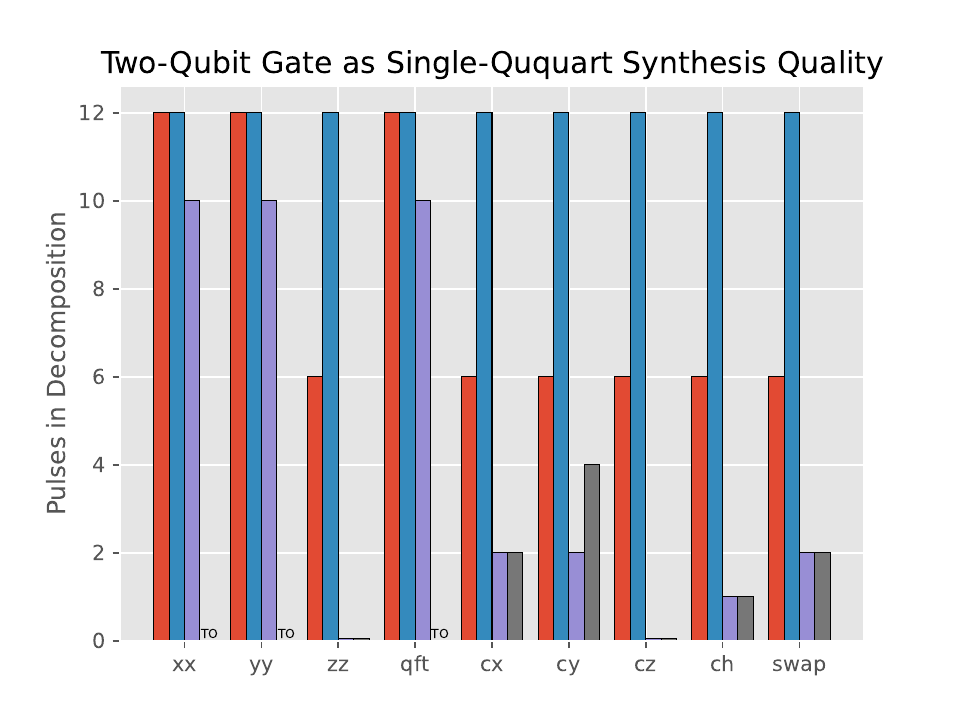}
    \includegraphics[width=\columnwidth]{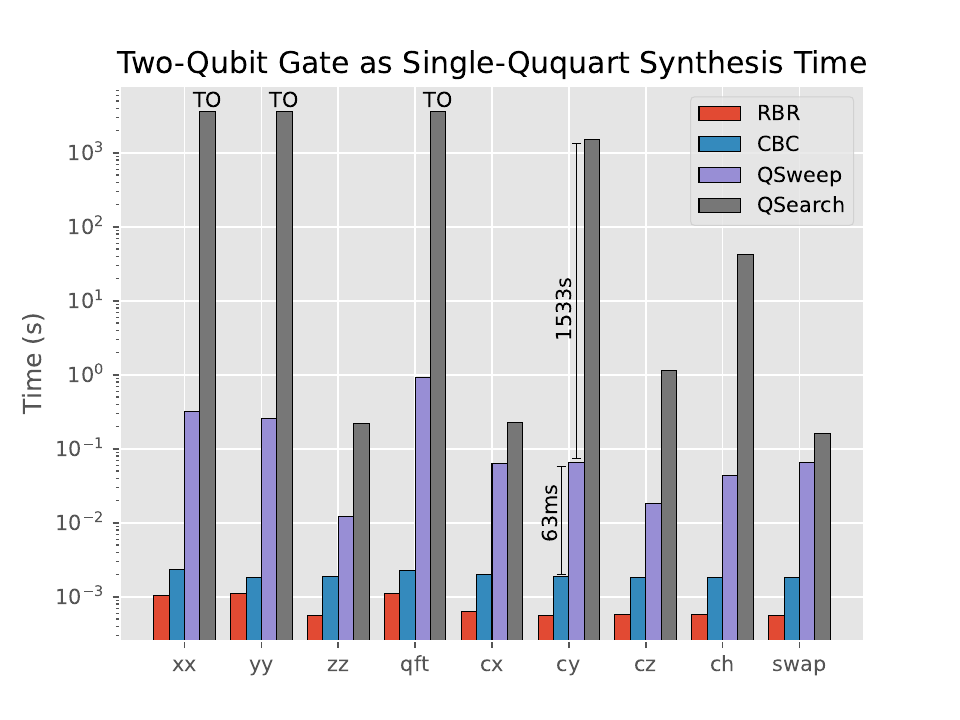}
    \caption{
        \footnotesize \it Synthesis of several standard two-qubit gates as single-ququart ones. Both quality (left) and time (right) are plotted for the four algorithms. QSearch timed out after an hour on three benchmarks denoted by a ``TO.''
    }
    \label{fig:2qgate}
\end{figure*}

\begin{figure}[t]
    \centering
    \includegraphics[width=\columnwidth]{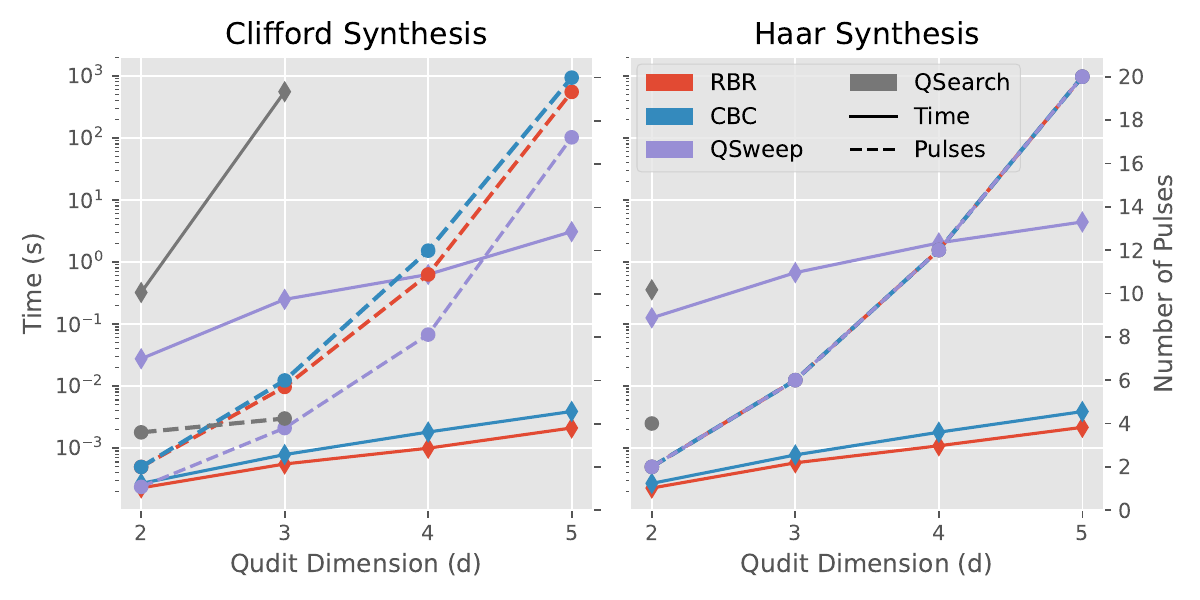}
    \caption{
        \footnotesize \it Random Clifford and Haar unitaries were synthesized for varying dimensions using different methods. Each data point is the average of 100 random decompositions.
    }
    \label{fig:random}
\end{figure}

\begin{figure}
    \centering
    \includegraphics[width = 0.85\columnwidth]{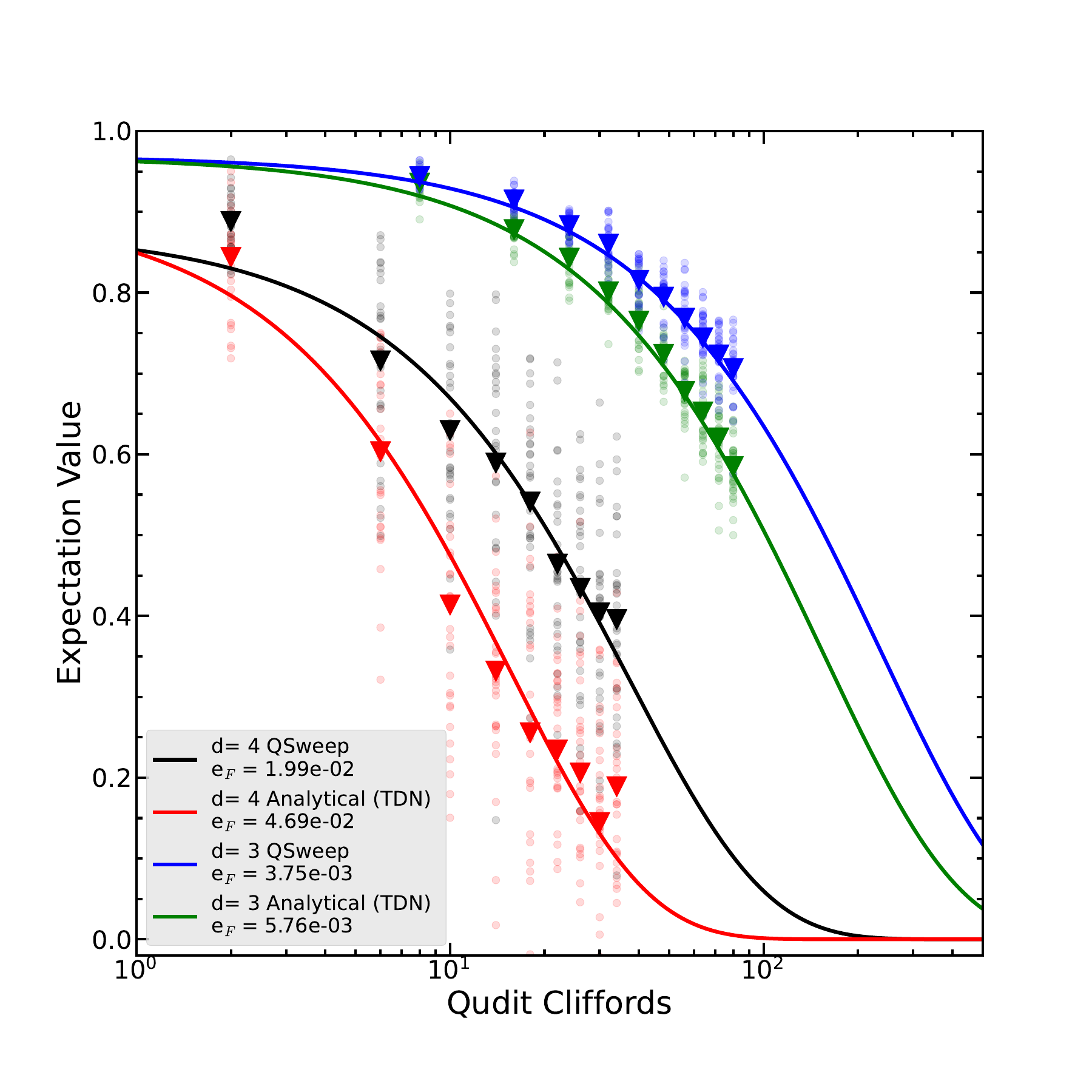}
    \caption{
        \footnotesize \it Qudit randomized benchmarking results for dimension $d=3$ and $d=4$ where each Clifford gate is decomposed with a top down (TDN) analtyical method and with QSweep.
    }
    \label{fig:RB}
\end{figure}

We evaluated four synthesis algorithms, two analytical and two numerical, on randomly generated Clifford and Haar-sampled unitaries with varying dimensions. The radix ranged from two (qubits) to five (ququints) with 100 random unitaries for each data point. Figure~\ref{fig:random} plots the time and quality for both types of unitaries. We did not run QSearch for Cliffords with $d \ge 4$ and Haar unitaries $\ge 3$ because of lengthy runtimes.

Haar-sampled unitaries are uniformly sampled. The chance of selecting an operation expressible in less than the maximum number of gates is practically zero as the unitary space is massive. Accordingly, all synthesis tools yielded the same quality, except for qsearch with qubits. The row-by-row (RBR) method was the fastest by a slim margin over column-by-column (CBC), with QSweep trailing behind. The scaling of QSweep starts to show around $d = 5$, where it takes about 3 seconds to complete each decomposition, while the analytical decompositions are still well under one second.

While the timings are similar for both unitary types, there is a more significant variance in the quality of Clifford decompositions. QSweep produced shorter sequences for every dimension tested, requiring an average of 2.44, 1.93, and 1.48 fewer pulses than CBC, RBR, and Qsearch, respectively.

One interesting observation we made during the initial development of QSweep is that we can construct circuits with fewer pulses for Haar-random unitaries if we allow a greater degree of error in our synthesis. As QSweep is easily configurable, if we set a high $10^{-3}$ error threshold, we can find results with fewer pulses than expected. We do not evaluate this here to ensure apples-to-apples comparisons, but this may be a valid use depending on the context.

\subsection{Compressed Two-Qubit Gates}

Two qubits are equivalent to one ququart, and some compilers use this fact in a compression scheme~\cite{litteken2023qompress}. We also leveraged this fact to build a benchmark set out of standard two-qubit gates converted to single-ququart gates since there is no established suite for ququart problems. These well-structured input unitaries add diversity to our evaluation beyond random benchmarks. Figure~\ref{fig:2qgate} depicts the results of the experiment.

CBC performed consistently across all benchmarks, producing 12-pulse circuits in about 1.95 milliseconds, and serves as the baseline. Our adaptive RBR analytical decomposition outperforms CBC, yielding six pulses in six of the nine trials while also being $2.6\times$ faster. Yet, QSweep synthesized the highest-quality circuits in every case, requiring an average of 7.9, 3.9 and 0.3 fewer pulses than CBC, RBR, and QSearch, respectively. While QSearch matched QSweep's quality in all but one case, it required an average of 4100x (up to 23500x) more runtime than QSweep. Furthermore, the CY benchmark highlights the differences between QSweep and QSearch's optimality criterion: QSweep prioritizes fewer pulses, while QSearch targets fewer gates.

\subsection{Experimental Demonstration}

We experimentally demonstrated the benefit of QSweep via randomized benchmarking on a superconducting quantum device capable of qudit computing. This machine's software stack was set to perform analytical decomposition based on a top-down algorithm, which produces the same circuits as the column-by-column decomposition in previous evaluations. Therefore, we compared our QSweep decomposed circuits to the top-down (TDN) method, resulting in a 1.54x and 2.36x improvement in single-qutrit and ququart gate fidelity.


\section{Discussion and Conclusion}
\label{conclusion}

We developed and demonstrated the QSweep algorithm, a numerical synthesizer that can generate pulse-optimal single-qudit decompositions for any given gateset. Although we significantly improved runtime performance compared to previous numerical synthesizers, we only created a prototype of our algorithm. We believe that with proper engineering and parallelization efforts, QSweep's runtimes can be further improved significantly in a production setting. This belief stems from the observation that all numerical optimization subroutines were extremely quick, typically finishing in a few steps.

Throughout our work, we operated under the assumption that a qudit's lower levels are more resistant to errors than its higher levels. This hypothesis did lead to a substantial improvement in experimental fidelity. However, ignoring the higher levels during most of a gate's execution could be problematic because they are also susceptible to dephasing. There may be an alternative elimination method that leads to a different interaction architecture with dynamical decoupling benefits, further improving fidelity.

\section*{Acknowledgment}
This work was supported by the U.S. DOE, Office of Science,
Office of Advanced Scientific Computing Research (ASCR)
under contract No. DE-AC02-05CH11231 through the Accelerated
Research in Quantum Computing (ARQC) program and the Testbeds
for Science program.


\bibliographystyle{IEEEtran}
\bibliography{quantum}

\end{document}